\documentclass[twocolumn,showpacs,preprintnumbers,amsmath,amssymb]{revtex4}

\usepackage{graphicx}
\usepackage{dcolumn}
\usepackage{bm}

\begin{document}

\preprint{APS/123-QED}

\title{Magnetization-induced optical third-harmonic generation in $Co$ and $Fe$ nanostructures}
\author{O.A. Aktsipetrov}%
 \email{aktsip@shg.ru}
\author{ E.M. Kim, R.V. Kapra}
\author{T.V. Murzina}%
 \email{mur@shg.ru}

\affiliation{%
Department of Physics, Moscow State University, 119992 Moscow,
Russia}%
\author{A.F. Kravets}
\affiliation{%
Institute of Magnetism, National Academy of Sciences of Ukraine, Kiev, 03680 Ukraine}%
\author{M. Inoue}
\affiliation{%
Toyohashi University of Technology, Toyohashi 441-8580 Japan
}%
\author{S.V. Kuznetsova, M.V. Ivanchenko, and V.G. Lifshits }
\affiliation{%
Institute of Automation and Control Processes, 690041 Vladivostok, Russia}%

\date{\today}

\begin{abstract}
{Magnetization-induced optical third-harmonic generation (MTHG) is
observed in magnetic nanostructures: $Co$ and $Fe$ nanolayers and
granular films containing $Co$ nanoparticles.
Magnetization-induced variations of the MTHG characteristics in
these nanostructures exceed the typical values of linear
magneto-optical Kerr effect by at least an order of magnitude: the
maximum of magnetic contrast in the MTHG intensity is up to 0.2,
the angle of polarization rotation for MTHG is $10^{\circ}$ and
the relative phase shift is up to $100^{\circ}$.}
\end{abstract}

\maketitle

Magneto-optics, with its more than century-long history, remains
one of the most important experimental methods in studies of
magnetism. Recently, significant attention has been directed
towards nonlinear magneto-optics [1]: nonlinear magneto-optical
Faraday and Kerr effects in second-harmonic generation (SHG) were
observed experimentally in yttrium-iron-garnet films [2], then at
surfaces of magnetic metals [3], in magnetic multilayers [4] and
nanogranules [5]. Experimental measurements [2-5] and theoretical
estimates [6] reveal the typical magnitudes of the
magnetization-induced effects in SHG: the magnetization-induced
variations of the SHG intensity and rotation of the
second-harmonic (SH) wave polarization, may exceed the linear
magneto-optical Kerr effect (MOKE) by orders of magnitude.

It is well recognized that magnetization-induced SHG (MSHG) is a
powerful probe of surface and interface magnetism [1,6,7]. The
latter comes about from the lack of dipole magnetization-induced
second-order susceptibility in the bulk of centrosymmetric
magnetic materials. In contrast to the bulk, the surface of
centrosymmetric materials is always noncentrosymmetric and
results in localization of the dipole magnetization-induced
nonlinear polarization at the surface. MSHG comes from the surface
layer that is several nanometers thick and probes the surface
magnetic properties. Meanwhile, this high surface sensitivity of
MSHG is extruded in shadow third-order nonlinear magneto-optical
effects because of their bulk localization as well as their small
expected value. This is why the observation of
magnetization-induced third-harmonic generation (MTHG) is
discussed just recently in garnet films [8,9] and magnetic alloys
[10]. On the other hand, both the surface MSHG and bulk MTHG are
expected to supplement each other as magneto-optical probes and
provide complementary information about the magnetic properties of
nanostructures.

In this paper, magnetization-induced THG is observed in the following magnetic
nanostructures: $Co$ and $Fe$ nanolayers, and granular films
containing $Co$ nanoparticles.

For the description of MTHG, one can follow the phenomenological
approach which has been developed for the MSHG effects [1].
According to this approach, the third-order susceptibility of a
magnetic material is a combination of crystallographic and
magnetic terms, which possess \textit{even} and \textit{odd}
parities in magnetization, $M$, respectively:
$\chi_{ijkl}^{(3)}=\chi_{ijkl}^{(3) cryst}+\chi_{ijkl}^{(3)
odd}(M)$, where
$\chi_{ijkl}^{(3)cryst}(M)$=$\chi_{ijkl}^{(3)cryst}(-M)$ is the
crystallographic nonmagnetic susceptibility and $\chi_{ijkl}^{(3)
odd}(M)$=$-\chi_{ijkl}^{(3) odd}(-M)$ is the magnetic
susceptibility. Then the MTHG intensity is given by:
\begin{equation}
\begin{array}{l}
I_{3\omega}(M)\sim[\textbf{E}_{3\omega}^{cryst}+\textbf{E}_{3\omega}^{odd}(M)]^{2}=
[f_{i}(3\omega)(\chi_{ijkl}^{(3)cryst}+\\\chi_{ijkl}^{(3)odd}(M))
f_{j}(\omega)f_{k}(\omega)f_{l}(\omega)E_{j}(\omega)E_{k}(\omega)E_{l}(\omega)+c.c.]^{2},\label{1}
\end{array}
\end{equation}
where $E_{i}(\omega)$ is the $i$-th component of the fundamental
field, $\textbf{E}_{3\omega}^{cryst}$ and
$\textbf{E}_{3\omega}^{odd}(M)$ are the third-harmonic (TH) fields
originating from the crystallographic and magnetic
susceptibilities, respectively, $f_{i}(3\omega)$ and
$f_{j}(\omega)$ are coefficients that contain Fresnel and local
field factors and linear magneto-optical rotation of polarization
of fields at corresponding wavelengths. Eq. 1 includes a
cross-product of $\textbf{E}_{3\omega}^{cryst}$ and
$\textbf{E}_{3\omega}^{odd}(M)$. This interference cross-product,
$\textbf{E}_{3\omega}^{odd}(M) \cdot
\textbf{E}_{3\omega}^{cryst}$, is \textit{odd} with respect to
magnetization and is responsible for the internal homodyne effect
[11]. The absolute value of homodyne cross-term is not supposed to
be small, even for an intrinsically small magnetic co-factor, because
of the large value of crystallographic co-factor. To describe the
intensity effects in MTHG, the THG magnetic contrast,
$\rho_{3\omega}$, can be defined by analogy with the SHG magnetic
contrast, $\rho_{2\omega}$, introduced in [3,4]:
\begin{equation}
\rho_{3\omega}=\frac{I_{3\omega}(M\uparrow)-I_{3\omega}(M\downarrow)}
{I_{3\omega}(M\uparrow)+I_{3\omega}(M\downarrow)}\sim
\frac{\chi_{eff}^{(3)odd}(M)}{\chi_{eff}^{(3)cryst}}
\cos[\Phi_{3\omega}(M)], \label{1}
\end{equation}
where $I_{3\omega}(M\uparrow \downarrow)$ are the THG intensities
for the opposite directions of the magnetization,
$\chi_{eff}^{(3)odd}(M)$ and $\chi_{eff}^{(3)cryst}$ are the
combinations of tensor components of the corresponding
susceptibilities and $\Phi_{3\omega}(M)$ is the relative phase
between $\textbf{E}_{3\omega}^{cryst}$ and
$\textbf{E}_{3\omega}^{odd}(M)$. Symmetry analysis of
$\chi_{ijkl}^{(3)odd}(M)$ tensor [8,9] shows that
magnetization-induced variations of the THG intensity result from
the internal homodyne effect in the geometry of the transversal
Kerr effect (see the left-hand inset in Figure 1, $\emph{a}$). To
extract the ratio of \textit{magnetic} and
\textit{crystallographic} components of ${\chi}_{eff}^{(3)}$ from
$\rho_{3\omega}$, one should obtain $\Phi_{3\omega}(M)$ from the
THG interferometry.

The THG interferometry is the direct analogue of the SHG
interferometry described in detail elsewhere [12]. The scheme of
the nonlinear optical interferometry is shown at the right-hand
inset in Fig. 1 a).  The TH fields from the sample and the
reference source interfere at the photomultiplier while the
reference THG source is translated along the direction of the
fundamental beam. The interference pattern, i.e., an oscillating
dependence of the detected THG intensity as a function of the
reference displacement, $r$, results from the phase shift between
the interfering TH fields. This shift appears in the space between
the sample and the reference due to the dispersion of air at the
TH and fundamental wavelengths. The total THG intensity from a
magnetic sample and a nonmagnetic reference as a function of $r$
is given by:
\begin{equation}
\begin{array}{l}
I_{3\omega}(r,M)=
I_{3\omega}^{ref}+I_{3\omega}^{samp}(M)+2\alpha \sqrt{I_{3\omega}^{ref}I_{3\omega}^{samp}(M)}\\
\cos [\displaystyle \frac{2\pi
r}{L(3\omega)}+\Phi_{3\omega}^{samp}(M)-\Phi_{3\omega}^{ref}],\label{3}
\end{array}
\end{equation}
where $L(3\omega)=\lambda_{\omega}(3\Delta n)^{-1}$ is the
interference pattern period, $\lambda_{\omega}$ is the fundamental
wavelength, $\alpha$ is the coherence coefficient of the
fundamental beam, $\Delta n=n(3\omega)-n(\omega)$ describes the
dispersion of the refractive index  of air, $n$, at the THG and
fundamental wavelengths, respectively,
$\Phi_{3\omega}^{samp}(M)\equiv arg[E_{3\omega}^{samp}(M)]$ and
$\Phi_{3\omega}^{ref}\equiv arg[E_{3\omega}^{ref}]$ are the phases
of the TH fields from the magnetic sample,
$\textbf{E}_{3\omega}^{samp}(M)$, and the reference THG source,
$\textbf{E}_{3\omega}^{ref}$, respectively.  The
magnetization-induced shift between two interference patterns
measured for the opposite directions of the magnetic field,
$\varphi_{3\omega}(M)=\Phi_{3\omega}^{samp}(M\uparrow)-\Phi_{3\omega}^{samp}(M\downarrow)$,
allows one to deduce $\Phi_{3\omega}(M)$ in accordance with the vector
diagram at the inset in Fig. 1, b).

The samples of magnetic nanostructures studied in this work are:
(1) magnetic nanogranular films of the composition
$Co_{x}Ag_{1-x}$ and $Co_{x}(Al_{2}O_{3})_{1-x}$ and (2) thin
homogeneous $Co$ and $Fe(110)$ films. The $Co_{x}Ag_{1-x}$ and
$Co_{x}(Al_{2}O_{3})_{1-x}$ films are prepared by the
co-evaporation of $Co$ and $Ag$ ($Al_{2}O_{3}$) from two
independent electron-beam sources onto glass-ceramic substrates
[13]. The structure of $Co_{x}Ag_{1-x}$ and
$Co_{x}(Al_{2}O_{3})_{1-x}$ films is characterized by X-ray
diffraction and reveals the existence of nanogranules with the
diameter ranging from 3 nm to 6 nm for the composition
$\emph{x}<0.4$. The fabricated granular films exhibit giant
magnetoresistance (GMR) effects that are characterized by the GMR
coefficient: $\displaystyle\rho_{GMR}=-[R(0)-R(M)]/R(0)$, where
$R(M)$ and $R(0)$ is Ohmic resistance measured in magnetized and
demagnetized material. $\rho_{GMR}$ is measured by the four-probe
method at room temperature in a magnetic field up to 8 kOe.

Homogeneous $Co$ films 400 nm thick are deposited in
similar conditions using a single \emph{Co} source. The
homogeneous $Fe$(110) films 100 nm thick are epitaxially grown on
$Si$(111) substrates [14].

The output of an OPO laser system "Spectra-Physics ÌÎÐÎ 710" at
800 nm wavelength, pulse duration of 4 ns, pulse intensity of 2
$MW/cm^2$, and a Q-switched $YAG:Nd^{3+}$ laser at 1064 nm
wavelength, pulse duration of 15 ns and pulse intensity of
1 $MW/cm^2$ are used as the fundamental radiation. The TH(SH)
radiation is filtered out by appropriate glass bandpass and
interference filters and is detected by a PMT and gated electronics.
To normalize the THG(SHG) signal over the OPO and $YAG:Nd^{3+}$
laser fluency and the spectral sensitivity of the optical
detection system, a reference channel is used with a Z-cut quartz
plate as a reference and a detection system identical to that in
the "sample" channel. The THG and SHG interferometry is performed
by translating a 30 nm thick indium-tin-oxide (ITO) film on a
glass substrate in the direction parallel to the laser beam.
An in-plane dc-magnetic field up to 2 kOe  is applied in nonlinear
magneto-optical measurements  at the magnetic samples by permanent
Fe-Nd magnets.

Prior to the studies of magnetization-induced effects in THG, $Co$
and $Fe$ nanostructures are characterized by the MSHG probe. Table
1, column 2, presents the SHG magnetic contrast for all samples.
The largest value of $\rho_{2\omega}$  is up to 0.45 in $Fe$(110)
homogeneous nanolayers.
\begin{table*}
    \begin{tabular}{r||l|l|l|c|l|l|l|}
                                             & $\rho_{3\omega}$    &   $\rho_{2\omega}$  & $\varphi_{3\omega}(M)$  &  $\varphi_{2\omega}(M)$  &  $\chi_{eff}^{(3)odd}(M)/\chi_{eff}^{(3)cryst}$ & $\chi_{eff}^{(2)odd}(M)/\chi_{eff}^{(2)cryst}$\\
      \hline
      \hline
       $Co$                                  & $0.09\pm0.03$   &  $0.32\pm0.05$ & $70^{0}\pm10^{0}$   &    $14^{0}\pm4^{0}$        & $0.55\pm0.1$  & $0.2\pm0.05$  \\

      \hline
      $Fe(110)$                              & $0.08\pm0.03$  & $0.45\pm0.05$ & $9^{0}\pm5^{0}$       & $13^{0}\pm4^{0}$      & $0.09\pm0.05$   &  $0.3\pm0.05$ \\

      \hline
      $Co_{0.31}Ag_{0.69} $                  & $0.09\pm0.03$ &$0.05\pm0.03$     & $17^{0}\pm5^{0}$    & $8^{0}\pm3^{0}$        & $0.16\pm0.05$   &   $0.1\pm0.05$\\

      \hline
      $Co_{0.19}(Al_{2}O_{3})_{0.81} $       & $0.16\pm0.04$  & $0.09\pm0.03$    & $105^{0}\pm15^{0}$  & $14^{0}\pm5^{0}$         & $1.3\pm0.1$   & $0.13\pm0.05$  \\
      \hline

\end{tabular}
\caption{ Magnetic contrast in the THG and SHG intensity; the
magnetization-induced phase shift  of the total TH,
$\varphi_{3\omega}(M)$, and SH, $\varphi_{2\omega}(M)$, fields;
the ratio of $\chi_{eff}^{(3)odd}(M)/ \chi_{eff}^{(3)cryst}$ and
$\chi_{eff}^{(2)odd}(M)/ \chi_{eff}^{(2)cryst}$ for magnetic
nanostructures.}
\end{table*}

The experimental values of $\rho_{3\omega}$ in $Co$ and $Fe$
nanostructures are presented in the Table 1, column 1. In
particular, $\rho_{3\omega}$ in $Co_{0.19}(Al_{2}O_{3})_{0.81}$ is
up to 0.16, which exceeds the linear MOKE by about two orders of
magnitude [15]. The comparison of $\rho_{3\omega}$ and
$\rho_{2\omega}$ in the same materials shows that for
nanogranular films the magnetic contrast in THG and SHG is nearly
the same, while in $Co$ and $Fe$ nanolayers $\rho_{2\omega}$ is
several times larger than $\rho_{3\omega}$.
\begin{figure}[!h]
\vspace{1 cm}
\begin{centering}
\includegraphics
[width=8.5cm]{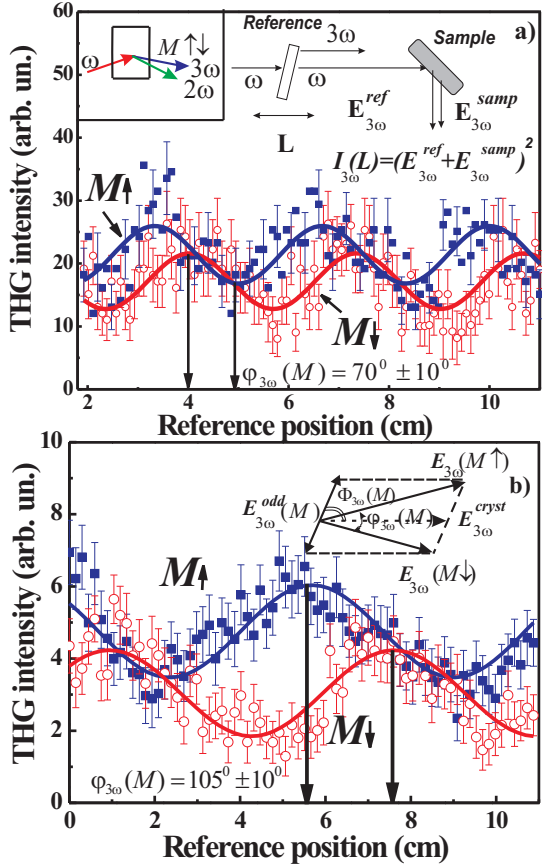} \caption{(a) The MTHG interferometry
patterns in  $Co$ film for the opposite directions of the
magnetization, $M\uparrow \downarrow$. Left-hand inset: the scheme of
the transversal geometry of nonlinear magneto-optical Kerr effect.
Right-hand inset: the scheme of SHG and THG interferometry. (b)
The MTHG interferometry patterns in
$Co_{0.19}(Al_{2}O_{3})_{0.81}$ film for the opposite directions
of the magnetization $M\uparrow \downarrow$. Inset: vector diagram
of odd and crystallographic components of the TH field.}\label{1}
\end{centering}
\vspace{0.1 cm}
\end{figure}

Figure 1 shows two THG interference patterns for $Co$ and
$Co_{0.19}(Al_{2}O_{3})_{0.81}$ films for the opposite directions
of magnetization and the magnetization-induced phase shift for
$\varphi_{3\omega}(M)=70^{0}$ and $105^{0}$, respectively, is
observed for the total TH field, $\textbf{E}_{3\omega}(M)$.
Results of interferometric measurements of the relative phase
shift of magnetization-induced TH field and the ratio of the
magnetic and crystallographic components of the third-order
susceptibility are summarized in Table 1.  Also shown are the results
of the analogous MSHG measurements which are presented for comparison with
second- and third-order nonlinear magneto-optical effects.

In the longitudinal NOMOKE configuration, the nonmagnetic and
magnetization-induced components of the TH field are polarized
orthogonally, $\textbf{E}_{3\omega}^{cryst}$ being {\it
p}-polarized and $\textbf{E}_{3\omega}^{odd}(M)$  being $s$-polarized,
respectively. The magnetization-induced effects appear in the rotation
of polarization of the total TH wave. The THG intensity depends on
the analyzer angle $\Theta$:
\begin{equation}
I_{3\omega}(\Theta)\propto|\textbf{E}_{3\omega}^{cryst}\cos\Theta+
\textbf{E}_{3\omega}^{odd}(M)\exp(i\phi_M)\sin\Theta|^2,
\end{equation}
where the phase shift $\phi_M$ describes the TH field ellipticity.
The TH wave is considered to be linearly polarized with
$\phi_M\simeq0$. The rotation angle of the polarization of the TH
wave upon reversal of magnetization is estimated to be
$\Delta\Theta\simeq2\arctan[|\textbf{E}_{3\omega}^{odd}(M)|/|\textbf{E}_{3\omega}^{cryst}|]$
and depends on the ratio of corresponding elements of the
$\chi_{ijkl}^{(3)odd}(M)$ and $\chi_{ijkl}^{(3)cryst}$ tensors
[10]. Magnetization-induced rotation of the TH wave polarization
up to $10^{0}\pm2^{0}$ is observed in \textit{Fe}(110) nanolayers.
\begin{figure}[!h]
\vspace{1 cm}
\begin{centering}
\includegraphics
[width=8.5cm]{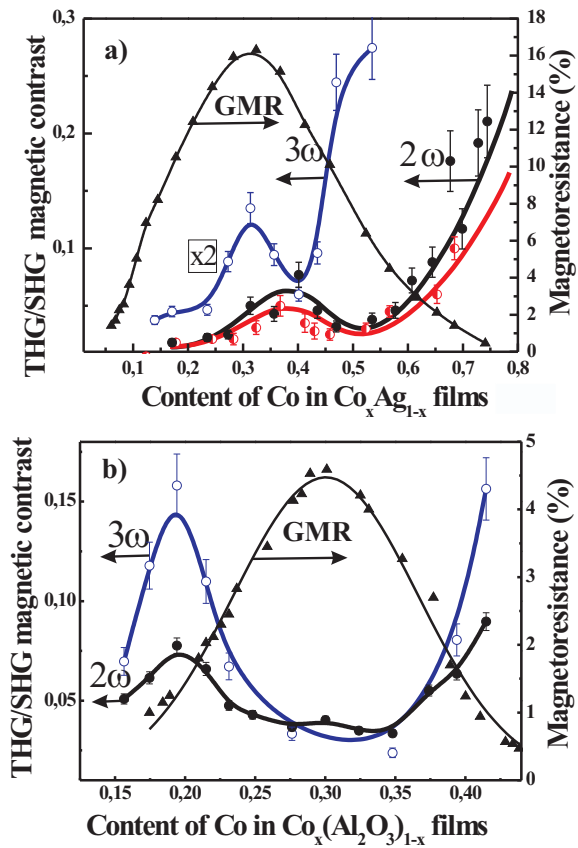} \caption {($\emph{a}$) Magnetic
contrast in the THG intensity for $Co_{x}Ag_{1-x}$ nanogranular
films as a function of the film composition $x$ (shown with open circles). The
SHG magnetic contrast at the fundamental wavelength of 1064nm (closed circles),
and 800nm (half-closed circles), respectively.
Magnetoresistance as a function of the $Co_{x}Ag_{1-x}$ film
composition $x$ (triangles). ($\emph{b}$) Magnetic contrast in the
THG intensity for $Co_{x}(Al_{2}O_{3})_{1-x}$ nanogranular films
as a function of the film composition $x$ (open circles). Magnetic
contrast in the SHG intensity (closed circles). The GMR coefficient
as a function of the $Co_{x}(Al_{2}O_{3})_{1-x}$ film composition
$x$ (triangles).}\label{1}
\end{centering}
\vspace{0.2 cm}
\end{figure}
The previously mentioned magnetization-induced effects in THG are attributed
to the proper combinations of components of magnetic and
crystallographic third-order susceptibilities. In principle, the
part of magnetization-induced changes in the THG parameters might
originate from linear MOKE at the fundamental wavelength. Meanwhile,
magnetization-induced changes of the amplitude and polarization of
the fundamental wave show that corresponding contributions to
MTHG are negligible.

The observation of magnetization-induced effects in THG from
$Co_{x}Ag_{1-x}$ and $Co_{x}(Al_{2}O_{3})_{1-x}$ nanogranular
films allows one to perform a comparative analysis of MTHG and GMR
effects as it was performed recently for MSHG in $Co_{x}Ag_{1-x}$
films [5]. Figure 2, $\emph{a}$ shows the dependence of
$\rho_{3\omega}$ on the concentration of $Co$ in $Co_{x}Ag_{1-x}$
films, which reveals a monotonic increase for $x>0.4$ and a local
maximum in the vicinity of $x\approx0.3$. The former corresponds
to a straightforward monotonic increase of the ferromagnetic phase
in the \textit{Co} fraction of the composite material as the
concentration of \textit{Co} exceeds the percolation threshold
[10]. A local maximum  corresponds to the specific magnetic
properties of nanogranules; $\rho_{3\omega}(x)$ attains a local
maximum at the same region of \textit{Co} concentration (0.3 -
0.35) as the concentration dependence of $\rho_{GMR}$. It is worth
noting that correlation between $\rho_{3\omega}(x)$ and
$\rho_{GMR}(x)$ is almost the same as for $\rho_{2\omega}(x)$ as
presented in Figure 2, $\emph{a}$ for comparison. The maximum of
$\rho_{2\omega}(x)$, measured at two wavelengths of the
fundamental radiation of 800 nm and 1064 nm, is attained in the
vicinity of $x\approx$0.36. The similarity of the dependencies of
$\rho_{2\omega}$ and $\rho_{GMR}$ on $x$ in $Co_{x}Ag_{1-x}$
nanogranular films has recently [5] been supposed to originate
from the dependence of the nonlinear optical response and the GMR
effect on the quality of interfaces between magnetic granules and
the nonmagnetic host material. In fact, GMR in $Co_{x}Ag_{1-x}$
nanogranular films is attributed to the spin-dependent electron
\textit{scattering} and is highly interface sensitive [13].

The connection between the GMR mechanism and the correlation between
$\rho_{3\omega/2\omega}(x)$ and $\rho_{GMR}(x)$ becomes more
apparent from the comparison of the MTHG and MSHG results in two
different types of nanogranular films. In contrast to
spin-dependent electron \textit{scattering} in $Co_{x}Ag_{1-x}$
films, the GMR effect in $Co_{x}(Al_{2}O_{3})_{1-x}$ films is
attributed to spin-dependent electron \textit{tunnelling}
[16]. Figure 2, $\emph{b}$ shows the dependencies of
$\rho_{3\omega/2\omega}$ on the concentration of $Co$ in
$Co_{x}(Al_{2}O_{3})_{1-x}$ films. Qualitatively, these dependencies
are close to those in $Co_{x}Ag_{1-x}$ films. Meanwhile, contrary
to $Co_{x}Ag_{1-x}$ films, the local maximum of both
$\rho_{3\omega}(x)$ and $\rho_{2\omega}(x)$ is attained at
approximately $x\approx0.19$, whereas the peak of $\rho_{GMR}(x)$
is centered at $x\approx0.3$. This shift between maxima of
$\rho_{3\omega/2\omega}(x)$ and $\rho_{GMR}(x)$ in
$Co_{x}(Al_{2}O_{3})_{1-x}$ films shows the lack of
correlation between nonlinear magneto-optical  and GMR effects in
this material.
\par In conclusion, magnetization-induced optical third-harmonic generation is
observed in magnetic homogeneous $Fe$ and $Co$ nanolayers and
magnetic films containing $Co$ nanogranules. Magnetization-induced
variations of the THG parameters are of the same order of
magnitude as in MSHG, and exceed the typical values of the linear
MOKE by at least an order of magnitude. Correlation between the
dependencies of $\rho_{3\omega/2\omega}(x)$ and $\rho_{GMR}(x)$ is
observed below the percolation threshold in $Co_{x}Ag_{1-x}$
films, which reveals the spin-dependent \textit{scattering} GMR
mechanism, whereas in $Co_{x}(Al_{2}O_{3})_{1-x}$ films with the
spin-dependent \textit{tunnelling} GMR mechanism, the dependencies
of $\rho_{3\omega/2\omega}(x)$ and $\rho_{GMR}(x)$  do not
correlate. At the same time, the general conclusion about
correlation between MTHG and GMR in $Co_{x}Ag_{1-x}$ films demands
further detailed studies.
\begin{acknowledgments}
This work is supported in part by the International Association
(INTAS) Grant No. 03-51-3784, the Grant-In-Aid from the Ministry
of Education, Science, Culture and Sport of Japan No. 14655119 and
the Presidential Grant for Leading Russian Science Schools No.
1604.2003.2.
\end{acknowledgments}

\end{document}